\documentclass[prl,twocolumn,superscriptaddress,floatfix,noshowpacs,10pt,longbibliography]{revtex4-2}

\usepackage{lipsum}

\usepackage{graphicx,bm,times}
\usepackage{amsmath}
\usepackage{amsfonts}
\usepackage{amssymb}
\usepackage{bm}% bold math
\usepackage{color}
\usepackage{times}
\usepackage{multirow}
\usepackage[%
  colorlinks=true,
  urlcolor=blue,
  linkcolor=blue,
  citecolor=blue
  ]{hyperref}
\DeclareUnicodeCharacter{2212}{-}
\DeclareUnicodeCharacter{2009}{\,}
\DeclareUnicodeCharacter{0301}{\'{\i}}

\usepackage{soul}
\usepackage{xcolor}

\graphicspath{{figures/}}

\begin{document}

 \title{{Drastic field-induced resistivity upturns 
as signatures of unconventional magnetism in superconducting iron chalcogenides}}

\author{Z. Zajicek}
\affiliation{Clarendon Laboratory, Department of Physics,
	University of Oxford, Parks Road, Oxford OX1 3PU, UK}

\author{I. Paulescu}
\affiliation{Clarendon Laboratory, Department of Physics,
	University of Oxford, Parks Road, Oxford OX1 3PU, UK}

\author{P. Reiss}
\affiliation{Clarendon Laboratory, Department of Physics,
	University of Oxford, Parks Road, Oxford OX1 3PU, UK}
\thanks{Current affiliation: Max Planck Institute for Solid State Research, Stuttgart, Germany}

\author{R. M. Abedin}
\affiliation{Clarendon Laboratory, Department of Physics,
	University of Oxford, Parks Road, Oxford OX1 3PU, UK}

\author{K. Sun}
\affiliation{Clarendon Laboratory, Department of Physics,
	University of Oxford, Parks Road, Oxford OX1 3PU, UK}
 
\author{S. J. Singh}
\thanks{Current affiliation: Institute of High-Pressure Physics, Polish Academy of Sciences, Sokolowska 29/37, 01-142, Warsaw, Poland}
\affiliation{Clarendon Laboratory, Department of Physics,
University of Oxford, Parks Road, Oxford OX1 3PU, UK}

\author{A. A. Haghighirad}
\affiliation{Clarendon Laboratory, Department of Physics,
	University of Oxford, Parks Road, Oxford OX1 3PU, UK}
\affiliation{Institute for Quantum Materials and Technologies, Karlsruhe Institute of Technology, 76021 Karlsruhe, Germany}

\author{A. I. Coldea}
\email[corresponding author:]{amalia.coldea@physics.ox.ac.uk}
\affiliation{Clarendon Laboratory, Department of Physics, University of Oxford, Parks Road, Oxford OX1 3PU, UK}

\begin{abstract}

Electronic scattering is a powerful tool to identify underlying
changes in electronic behavior and incipient electronic and magnetic orders.
The nematic and magnetic phases are strongly intertwined under applied pressure in FeSe,
however, the additional isoelectronic substitution of sulphur offers an elegant way to separate them.
Here we report the detailed evolution of the electronic and superconducting behaviour of FeSe$_{0.96}$S$_{0.04}$ under applied pressure via longitudinal magnetoresistance studies up to 15~T.
At intermediate pressures, inside the nematic phase,
the resistivity displays an upturn in zero magnetic field,
which is significantly enhanced in the magnetic field, suggesting the
stabilization of a spin-density wave phase, which competes with superconductivity.
At higher pressures, beyond the nematic phase boundaries, the resistivity no longer displays any clear anomalies in the zero magnetic field, but an external magnetic field
induces significant upturns in resistivity reflecting a field-induced order,  where superconductivity and magnetic anomalies are enhanced in tandem.
This study
highlights the essential role of
high magnetic fields in
stabilizing different electronic phases and revealing a complex interplay between magnetism and superconductivity
tuned by applied pressure in FeSe$_{1-x}$S$_{x}$. 
\end{abstract}
\date{\today}
\maketitle

The phase diagrams of unconventional high-$T_{\rm c}$ superconductors are often complex with multiple competing electronic phases \cite{Fernandes2022}.
High magnetic fields at low temperatures are extremely important for understanding the interplay of these phases, as they can suppress the superconducting phases and reveal the underlying electronic states from which they emerge and access Fermi surfaces via quantum oscillations \cite{Zajicek2024}. Resistivity upturns under applied magnetic fields at low temperatures are linked to signatures of potential competing orders with superconductivity \cite{Chen2009}. One prominent mechanism that gives such  upturns in electronic transport is the weak localization effect, where quantum interference enhances backscattering due to disorder \cite{bergmann1984}
or the Kondo effect, where conduction electrons scatter off localized magnetic moments
\cite{kondo1964}.
Additionally, the presence of charge or spin-density wave  
may lead to resistivity upturns due to partial gapping of the Fermi surfaces \cite{gruner1988,Zhang2025,Terashima2014}.
On the other hand, magnetic freezing effects from disordered magnetic moments can enhance resistivity, such as in spin glasses \cite{mydosh1993}, whereas the formation of short-range magnetic order also lead to resistivity upturns, particularly in strongly correlated systems \cite{dagotto2001} or cuprate superconductors \cite{bourgeoishope2019,Boebinger1996}

The pressure-induced high-$T_{\rm c}$ superconductivity of FeSe is a particular example where intertwined
nematic electronic order and magnetism  are particularly relevant for stabilizing superconductivity
\cite{Mizuguchi2008,Medvedev2009,Sun2016pressure}.
At ambient pressure, FeSe has a nematic electronic phase, induced by orbitally-dependent electronic correlations, and a highly anisotropic superconducting phase is found below 9\,K \cite{Hsu2008,Sprau2017}. 
Despite the absence of any long-range magnetic order in FeSe, there is a competition between N{\'e}el and stripe spin fluctuations \cite{Wang2016b}.
Applied pressure suppresses the nematic phase of FeSe and stabilizes a complex magnetic order \cite{Gati2019}. Superconductivity is significantly enhanced, reaching a maximum $T_{\rm c}$ of 37\,K at high pressures
and spin fluctuations are proposed to mediate the superconducting pairing
~\cite{Medvedev2009,Sun2016pressure}.

The electronic nematic phase of FeSe is highly sensitive to both chemical substitution in the Fe plane, such as the substitution of Cu, and also out of the plane at the chalcogen position with the isovalent substitution of sulphur or telurium on selenium sites
\cite{Coldea2019,Mizuguchi2009,Mukasa2021,Zajicek2022}. The nematic phase of FeSe$_{1-x}$S$_{x}$  is suppressed beyond $x = 0.18$, while superconductivity develops a small dome inside it
and suffers significant changes in the gap symmetry inside the tetragonal phase \cite{Hanaguri2018,Coldea2019}.
The phase diagram of FeSe$_{1-x}$S$_{x}$ can be altered by combining applied and chemical pressure, by continuously suppressing the nematic phases and stabilizing enhanced superconductivity at high pressures \cite{Matsuura2017,Reiss2024}. In these systems, anomalies in resistivity, associated with a spin density wave (SDW), occur over a broad regime under pressure in FeSe ($\sim$8 to 63\,kbar) \cite{Sun2016pressure}, but with increasing $x$ this magnetic region narrows down (centered around 50\,kbar for $x \sim 0.11$) \cite{Matsuura2017}. 
This may indicate that the nature of the magnetic order in FeSe$_{1-x}$S$_{x}$ at low pressure,  inside the nematic phase \cite{Xiang2017}, may be distinct from that found at higher pressure in the tetragonal phase \cite{Matsuura2017}.
Thus, studying these systems in a detailed manner under pressure and high magnetic fields offers a unique insight into the role
of isoelectronic substitution in affecting different competing interactions
that influence superconductivity.

\begin{figure}[htbp]
	\centering
	\includegraphics[ width=1\linewidth,clip=true]
    {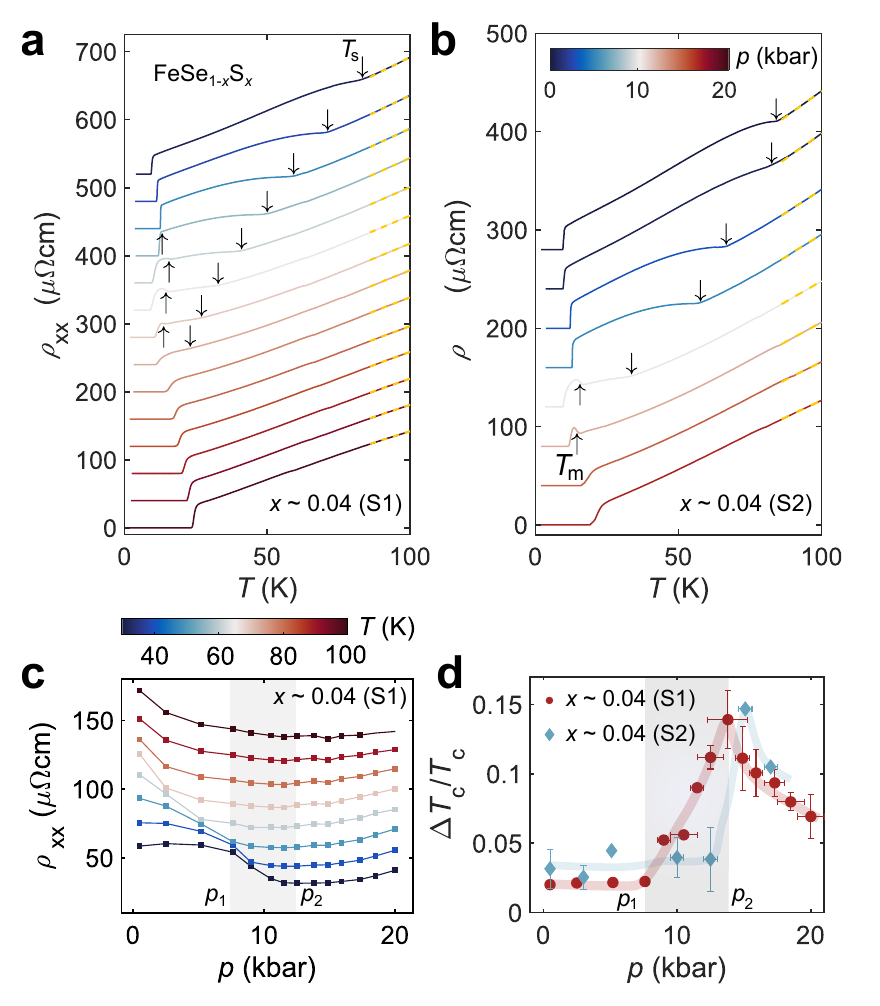}
\caption{{\bf The evolution of the transport behaviour under pressure in FeSe$_{0.96}$S$_{0.04}$.}
\textbf{(a)-(b)} Longitudinal resistivity $\rho_{xx}$ against temperature tuned under different applied pressure for  samples S1 and S2. The down arrows mark the nematic transition temperature at $T_{\rm s}$, whereas the up arrows indicate the minimum in the first-order derivative at $T_{\rm{m}}$.
The curves are shifted vertically for easier visualization.
\textbf{(c)} 
The pressure dependence of the resistivity at different fixed temperatures. 
Shaded gray areas indicate boundaries between different electronic phases at $p_1$ and $p_2$.
{\bf (d)} The relative changes in the superconducting transition width, $\Delta T_{\rm c}/T_{\rm c} = (T_{\rm c,mid} - T_{\rm c, off})/T_{\rm c, mid}$,
where $T_{\rm c,mid}$ is defined as the peak in first derivative (see 
Fig.~S1
in the SM \cite{SM}) for sample S1
(solid circles), S2 (solid diamonds), Fe$_{1-z}$Cu$_z$Se, $z=0.0025$ (after Ref.~\cite{Zajicek2022})
versus applied pressure.
Shaded gray areas indicate phase boundaries between different electronic
phases at $p_1$ and $p_2$.
}
	\label{Fig1_x4_zero_field}
\end{figure}

In this paper, we present a detailed transport study of FeSe$_{0.96}$S$_{0.04}$ under applied pressure 
to understand  the stabilization of field-induced electronic phases and their interplay with superconductivity.
Under applied pressure, the resistivity develops a sharp upturn in zero magnetic field,
which reflects the stabilization of a spin-density wave inside the nematic phase. At higher pressures, the upturn disappears and, instead, the transition from the normal to the superconducting phase becomes broader. 
However, the presence of a magnetic field induces a large magnetoresistance, and the upturn in resistivity persists over the whole pressure range, indicating the development of
field-induced electronic phases.
The phase diagram suggests that superconductivity has two different domes
that potentially reflect magnetically mediated pairing originating from different magnetic phases.

\paragraph{\bf Experimental Methods.}
Single crystals of FeSe$_{0.96}$S$_{0.04}$ were prepared using the chemical vapour transport method with KCl/AlCl$_{3}$ \cite{Chareev2013,Bohmer2016}. 
High-quality single crystals, considered for their sharp superconducting transitions and high $RRR$ values. Experiments were carried out on different samples (S1 and S2) from the same batch using a commercially available BeCu pressure cell from Quantum Design up to 20~kbar and cooling down slowly to 2~K at a rate of 0.5~K/min, inside a  16\,T PPMS (Physical property measurement system). A maximum a.c. current of 1\,mA (RMS) was applied to the sample. Daphne 7373 was used as the pressurising medium, which is hydrostatic up to 22~kbar \cite{Yokogawa2007}. The pressure was determined in situ using the superconducting transition temperature of tin at a slow cooling rate of 0.02\,K/min.

\begin{figure*}[htbp]
	\centering
	 \includegraphics[trim={0cm 0cm 0cm 0cm}, width=0.9\linewidth,clip=true]
    {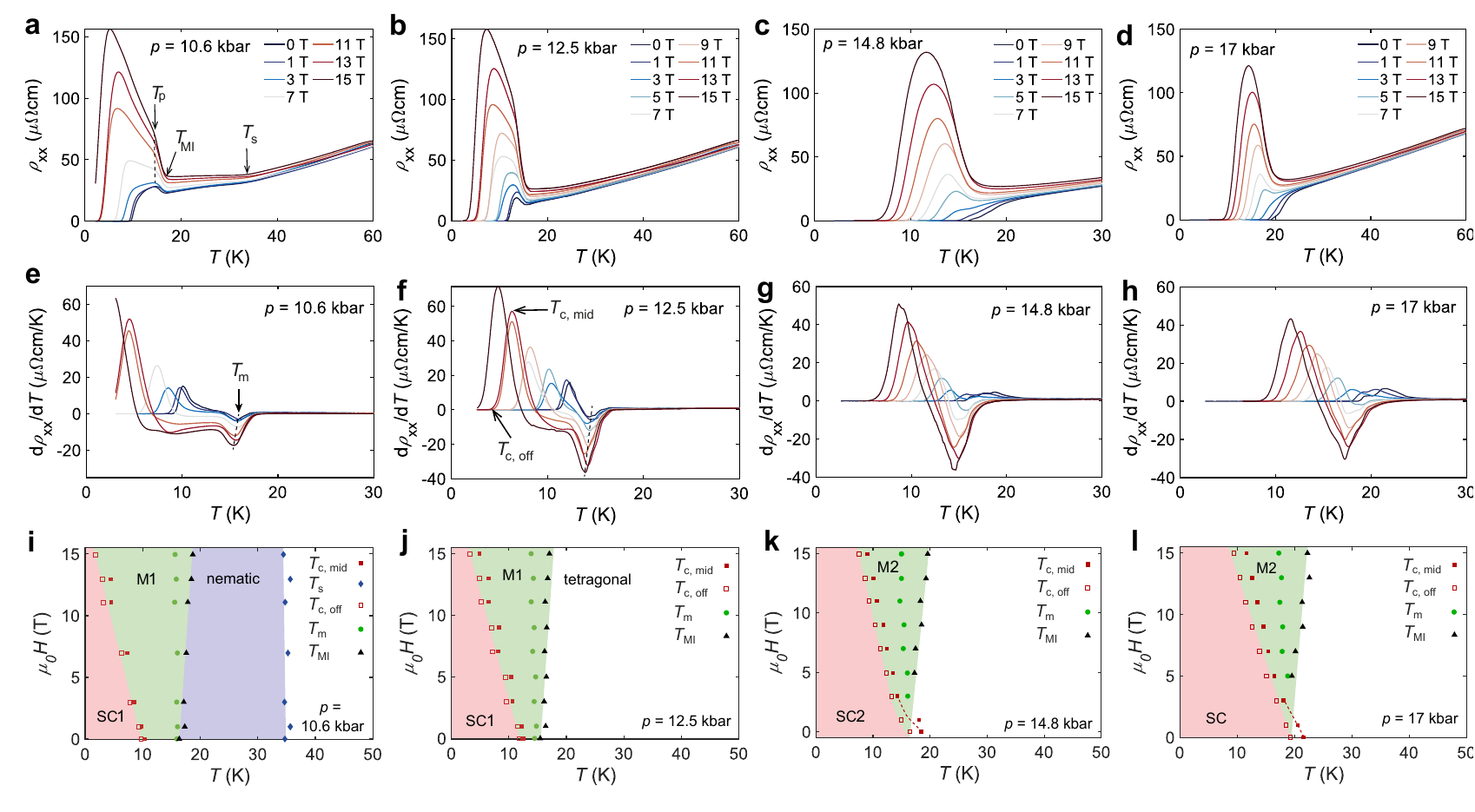}
\caption{{\bf The  magnetoresistance of FeSe$_{0.96}$S$_{0.04}$  for sample S2 tuned by applied pressure.}
{\bf (a) - (d)}.
Temperature dependence of the longitudinal resistivity in magnetic fields of up to 15~T under pressures equal or greater than $p_1 \sim 10.6$~kbar.
{\bf (e) - (h)} The corresponding first derivatives of $\rho_{\rm xx}$  with respect to temperature related to raw data in panels {\bf (a) - (d)}
The development of the new field-induced phase are defined as the minimum in the derivative at $T_{\rm m}$ and the local minimum in resistivity data at $T_{\rm MI}$, whereas critical temperature is defined here as the maximum in the derivative at the midpoint transition, $T_{\rm c, mid}$ (see 
Figs.~S1 and S2 in the SM \cite{SM}).
The magnetic field-temperature phase diagram indicating the different electronic phases, nematic (purple), magnetic M1 and M2 (green), superconducting SC1 and SC2 (red). The dashed lines close to the superconducting transition in zero magnetic field indicate the broadening of the superconducting transition widths close to $p_2$.
}
\label{Fig2_x4_TZ2_rhoxx_vs_T_fixB_fixp}
\end{figure*}

\paragraph{\bf Transport behaviour under pressure in zero-magnetic field.}
Figs.~\ref{Fig1_x4_zero_field}(a) and (b) show the evolution of resistivity with applied pressure for FeSe$_{0.96}$S$_{0.04}$ for sample S1
and sample S2, respectively.
At ambient pressure, the resistivity displays a metallic-like behavior until the system enters the superconducting state below $T_{\rm c} = $ 9.3\,K.  A kink in resistivity at $T_{\rm s}$ = 80(2)~K for sample S1 defines the onset of the nematic electronic phase, 
which is also accompanied by a tetragonal-orthorhombic transition. 
By applying pressure up to $p_1 =7.5$~ kbar, the resistivity is continuously suppressed inside the nematic phase (see Fig.~\ref{Fig1_x4_zero_field}(c)), but its resistivity variation as a function of temperature is much weaker due to the potential scattering of nematic domains.
At higher pressures inside the nematic phase for $T_{\rm s}<40$~K, the resistivity develops an upturn at $T_{MI}$ and a local maximum at $T_{p}$ in the absence of the magnetic field (see also Fig.~\ref{Fig2_x4_TZ2_rhoxx_vs_T_fixB_fixp}(a)).
Resistivity upturns in zero-magnetic field were previously detected in interplane resistance, $\rho_{zz}$ for $x$ = 0.043 \cite{Xiang2017}
and a thin 2.3 $\rm{\mu m }$ flake \cite{Xie2021}. 
Such feature  was associated with a magnetic phase transition driven by a spin-density wave (SDW) in FeSe \cite{Sun2016pressure}.
Once the nematic phase is suppressed above $p_2 = 12.5$\,kbar, the resistivity upturn is washed out while the superconducting transition width becomes broader (see Figs.~\ref{Fig1_x4_zero_field}(a), (b) and (f)).

To understand the trends in scattering under pressure, one can assess the changes in resistivity at constant temperature, as
shown in Fig.~\ref{Fig1_x4_zero_field}(c).
At high temperatures, the resistivity decreases smoothly with applied pressure, as expected, due to an increase in the electronic bandwidth, as shown in Fig.~\ref{Fig1_x4_zero_field}(c).
At temperatures below 100~K, the resistivity decreases with pressure inside the nematic phase, below $p_1$, but then increases in the tetragonal phase above $p_2$, suggesting significant changes in scattering, electronic correlations, or enhanced spin fluctuations (see Fig.~\ref{Fig1_x4_zero_field}(c)).
Interestingly, the temperature dependence of the resistivity in the high-pressure regime is  rather linear and broadly similar to other FeSe$_{1-x}$S$_x$ \cite{Xiang2017,Reiss2020}, Fe$_{1-z}$Cu$_z$Se \cite{Zajicek2022Cupressure}, and FeSe \cite{Sun2016pressure}, suggesting that the high-pressure phase is  a different electronic phase with higher resistivity.

The width of the superconducting transition changes significantly with pressure, as shown in Fig.~\ref{Fig1_x4_zero_field}(f). In the low pressure regime, $p < p_{\rm 1}$, the superconducting transitions are sharp ( $\Delta T_{\rm c} \sim 0.5$\,K). However, in the intermediate pressure regime, where signatures of magnetic order are found, the superconducting transition broadens significantly (close to 3~K, similar to previous studies
in Cu-FeSe \cite{Zajicek2022Cupressure}. 
Relative changes in the superconducting transition, $\Delta T_{\rm c}/T_{\rm c}$, have a significant enhancement just outside the boundaries of the nematic phase close to $p_2$, which
could suggest the presence of strong critical nematic or spin fluctuations \cite{Reiss2021,Gati2019}, the existence of quantum Griffiths phases \cite{Reiss2021} or the  manifestation of different competing electronic phases at high pressure.

\paragraph{\bf Magnetotransport behaviour under applied pressure}
High magnetic fields are used to suppress the superconducting phase and reveal anomalous transport behaviour at low temperatures.
Fig.~\ref{Fig2_x4_TZ2_rhoxx_vs_T_fixB_fixp}  describes the evolution with pressure of the temperature dependence of resistivity in different magnetic fields.
At intermediate pressures ($p_1<p<p_2$),
the resistivity develops a metallic-to-insulating transition at $T_{MI}$ and  an additional slope change  at $T_p$ in magnetic field, coinciding with the zero-field resistance peak, as shown in Fig.~\ref{Fig2_x4_TZ2_rhoxx_vs_T_fixB_fixp}(a) and (b).
The change in magnetotoresistance in 15~T is rather weak inside the nematic phase below $T_{\rm s}$ (a factor of 1.5 at 20~K) before increasing significantly below $T_{MI}$, by a factor of 2.75 at 14~K. This suggests that the magnetic field has a significant influence on the low-temperature electronic phase.

The field-induced resistivity upturns at low temperatures, once superconductivity is suppressed, closely resemble those previously observed in FeSe \cite{Terashima2016_MR} and suggest the stabilization of an SDW phase.
Field-induced SDWs are often found
in systems close to nesting instabilities, such as low-dimensional organics \cite{Kornilov2002,KorinHamzic1999}.
Below $T_{MI}$, resistivity has an activated-like behaviour
as a result of the opening of the SDW energy gap.
A strong magnetoelastic coupling 
could be responsible for the change in slope at $T_{p}$, similar to FeSe,  which displays an additional increase of the in-plane lattice parameters around 15~kbar inside the nematic phase \cite{Kothapalli2016}.
Interestingly, at higher pressures, $p>p_2$, despite the lack of any upturns in zero-magnetic field with much broader superconducting transitions (see Fig.~\ref{Fig1_x4_zero_field}(d)), we detect significant resistivity upturns induced by magnetic fields, similar to those observed at lower pressures. 
This reflects that a similar underlying  mechanism is responsible for the large resistivity upturns and the corresponding significant magnetoresistance in these systems
(see Figs.~\ref{Fig2_x4_TZ2_rhoxx_vs_T_fixB_fixp}(c) and (d)
and the corresponding derivatives in Figs.~\ref{Fig2_x4_TZ2_rhoxx_vs_T_fixB_fixp}(g) and (h)).

\begin{figure*}[htbp]
	\centering
	\includegraphics[trim={0cm 0cm 0cm 0cm}, width=1\linewidth,clip=true]{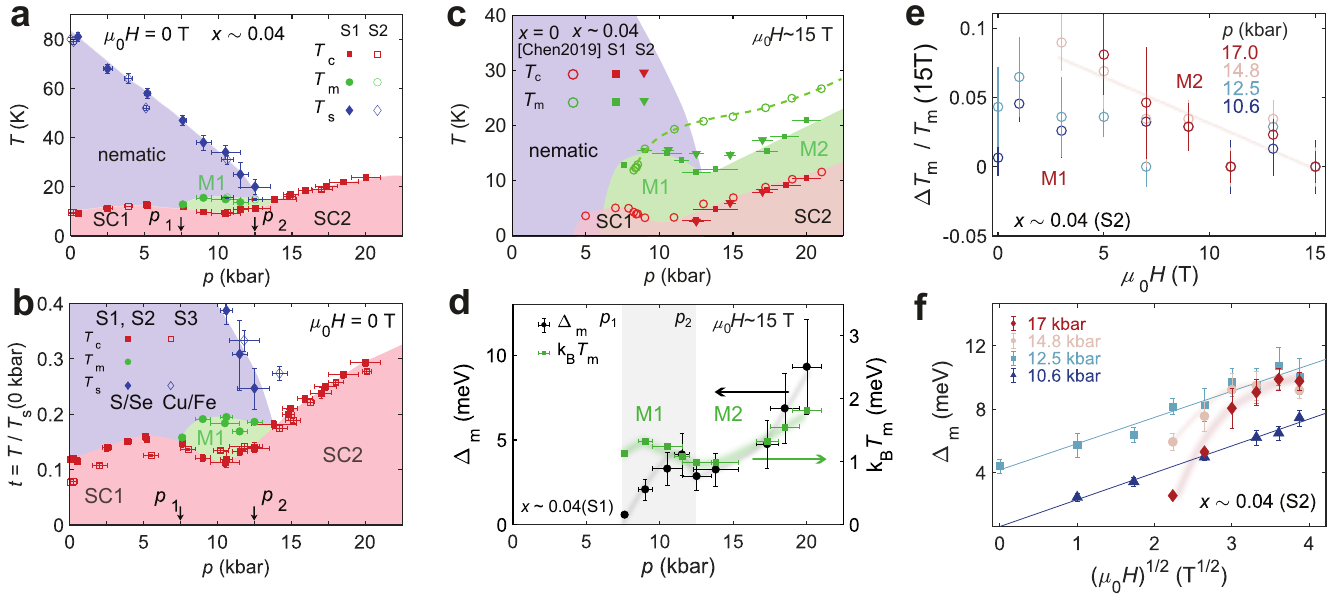}
\caption{{\bf Comparison of the $p-T$  phase diagrams and field-induced effects of FeSe$_{0.96}$S$_{0.04}$ under applied pressure.}
 {\bf (a)} Pressure-temperature phase diagram of FeSe$_{0.96}$S$_{0.04}$ for sample S1 (solid symbols) and sample S2 (open symbols) indicating
 the different competing electronic phases: nematic up to $p_1$ (purple area), magnetic (M1 green area) ($p_1<p<p_2$) and superconducting (SC1 and SC2).
 {\bf (b)}
The phase diagram of FeSe$_{0.96}$S$_{0.04}$ and Fe$_{0.9975}$Cu$_{0.0025}$Se in reduced temperature units in relation to their $T_{\rm s}$, $t=T/T_{\rm s}$.
 {\bf (c)}
 Pressure-temperature phase diagram of FeSe and FeSe$_{0.96}$S$_{0.04}$ in magnetic fields. 
 Symbols correspond to FeSe$_{0.96}$S$_{0.04}$ for samples S1 (solid green squares in 16~T) and S2 (solid triangles in 14~T) and FeSe  (open circles in  16\,T) after Ref.~\cite{GuanYu2019}. The green shaded region represents the magnetic phases in field (M1 and M2). 
 {\bf (d)} The pressure dependence of the effective energy, $\Delta_M$ (on the left $y$ axis) together with 
$k_{\rm B} T_{\rm m}$ in 15~T (on the right $y$ axis).
$\Delta_m$  is defined from the slope from $\log \rho$ versus $ 1/T$ in  
(see Fig.~S3 for S1 
and Fig.~S4 for S2 
in the SM \cite{SM}). 
{\bf (e)} The field dependence of the relative changes in the magnetic transition, $\Delta T_{\rm m}/T_{\rm m} = [T_{\rm m}(\mu_0 H) - T_{\rm m}({\rm 15~T})]/T_{\rm}$(15~T),
where $T_{\rm m}$ is taken as the minimum in the first derivative 
(see Fig.~\ref{Fig2_x4_TZ2_rhoxx_vs_T_fixB_fixp}({\bf e}-{\bf h})).  
{\bf (f)} 
The variation of the effective activation energy, $\Delta_m$,
as a function of square root of magnetic field, $H^{0.5}$,
for different pressures for sample S2 (see Fig.~\ref{Fig2_x4_TZ2_rhoxx_vs_T_fixB_fixp}(a-d)
and Fig.~S5 in the SM \cite{SM}).
The solid line is a linear fit to the data and the curved lines are guides to the eye.
}
\label{Fig_x4_PhaseDiagramsDiscussion}
\end{figure*}

The evolution of magnetic field-induced transitions at each pressure is quantified using
the sharp minimum at $T_m$ in the first derivative of
resistivity as a function of temperature (see Figs.~\ref{Fig2_x4_TZ2_rhoxx_vs_T_fixB_fixp}(e)-(h)).
This characteristic temperature, $T_{\rm m}$,
 decreases with increasing magnetic field as superconductivity is suppressed (see Figs.~\ref{Fig2_x4_TZ2_rhoxx_vs_T_fixB_fixp}(i)-(l)).
At intermediate pressure, the relative change in transition temperature, $\Delta T_{\rm m}/T_{\rm m }(15~{\rm T})$
slightly decreases
as the magnetic field increases (see Figs.~\ref{Fig2_x4_TZ2_rhoxx_vs_T_fixB_fixp}(e)-(h) and Fig.~\ref{Fig_x4_PhaseDiagramsDiscussion}(e)).
On the other hand,
at high pressures ($p>p_2$), 
the position of $T_{\rm m}$ is much more sensitive to magnetic fields, as shown in Fig.~\ref{Fig_x4_PhaseDiagramsDiscussion}(e).

At low temperatures and high magnetic fields, the resistivity increases exponentially as $\rho \sim \exp({\Delta_m/k_{\rm B} T_m})$ below $T_{MI}$, and the effective activation energy, $\Delta_{m}$,  can be estimated from the slope of the resistivity against the inverse temperature (see Figs.~S3, S4 and S5 in the SM \cite{SM}).
As a function of applied pressure, $\Delta_{m}$ closely follows the dependence of $T_{\rm m}$, as shown in Fig.~\ref{Fig_x4_PhaseDiagramsDiscussion}(d).
To understand how magnetic field induces changes in the resistivity, we follow the evolution of $\Delta_{m}$  for sample S2 shown in Fig.~\ref{Fig2_x4_TZ2_rhoxx_vs_T_fixB_fixp}.
We find that $\Delta_{m}$ increases with applied pressure phase and it has a field dependence of the form $H^{0.5}$ with a similar slope inside the nematic phase ($p_1<p<p_2$),
as shown in Fig.~\ref{Fig_x4_PhaseDiagramsDiscussion}(f).
On the other hand, at high pressures, $p>p_2$, $\Delta_{\rm m}$ is induced by magnetic fields and varies faster  towards saturation, indicating 
a strong sensitivity to field
(see Fig.~S5 in the SM \cite{SM}).
As $\Delta_{m}$ increases in magnetic field, the  $T_{MI}$ transition temperature shifts to higher values, as represented in the temperature-magnetic field phase diagrams at different pressures in Figs.~\ref{Fig2_x4_TZ2_rhoxx_vs_T_fixB_fixp}(i-l).
The dependence of $\Delta_{m}$
shows similarities with the field-induced magnetism detected in  underdoped La$_{2-x}$Sr$_x$CuO$_4$ ($x$ = 0.10),
where superconductivity and antiferromagnetism coexist \cite{Lake2002}.

When the magnetic order is accompanied by a tetragonal-orthorhombic transition, a first-order phase transition with hysteretic behaviour is expected, as found in FeSe at high pressure \cite{Bohmer2019,Kothapalli2016,Gati2019,Wang2016pressure}. 
We detect
just a weak 
hysteresis at 12.5~kbar between the cooling and warming curves 
(using a rate of 0.5~K/min for sample S2),
due to the magnetoelastic coupling at $T_p$ inside the nematic phase.
Therefore,  the isoelectronic substitution of sulphur for selenium in FeSe likely suppresses the strength of interactions and reduces the coupling between the magnetic and orthorhombic distortion (see  Fig.~S6 in the SM \cite{SM}).

\paragraph{\bf Discussion.}
Fig.~\ref{Fig_x4_PhaseDiagramsDiscussion}(a) shows the pressure-temperature phase diagram for different single crystals of FeSe$_{0.96}$S$_{0.04}$ based on the zero-field resistivity data.
This phase is split into three different regimes: the low-pressure nematic regime below $p < p_{\rm 1}$, the intermediate pressure regime, $p_{\rm 1} < p < p_{\rm 2}$, where the resistivity has an upturn at $T_{\rm m}$ in the absence of the magnetic field inside the nematic phase, and the high-pressure regime for $p > p_{\rm2}$, outside the nematic phase, where the resistivity increases and becomes progressively linear. 
The superconductivity displays a two-dome structure as a function of pressure, and at intermediate pressures $T_{\rm{c}}$ decreases whereas $T_{\rm m}$ increases, suggesting a competition between the two phases. 
Even for systems with higher sulphur substitution, like in FeSe$_{0.89}$S$_{0.11}$ under pressure, two similar superconductivity domes emerge, without having any clear resistivity upturns in a zero magnetic field \cite{Reiss2020}.

Fig.~\ref{Fig_x4_PhaseDiagramsDiscussion}(b) compares the zero-field phase diagram under pressure of FeSe$_{0.96}$S$_{0.04}$  for two different samples, S1 and S2, 
with that of Fe$_{1-z}$Cu$_z$Se ($z$ = 0.0025) previously reported in Ref.~\onlinecite{Zajicek2022Cupressure},
both scaled by the nematic transition $T_{\rm s}$ at ambient pressure.
At ambient pressure, sulphur substitution slightly enhances $T_{\rm c}$, compared to FeSe, while Cu substitution suppresses $T_{\rm c}$ \cite{Zajicek2022,Zajicek2022Cupressure}. 
Importantly,  the pressure variation of the
nematic and superconducting transitions is rather similar for the two systems.
However, in the intermediate pressure regime, the dome of magnetic order of FeSe$_{0.96}$S$_{0.04}$ is absent in Fe$_{1-z}$Cu$_z$Se, 
indicating that the increased impurity scattering potentially obscures the resistivity upturns in the absence of a magnetic field.
Interestingly, in the high-pressure phase above $p_2 = 12.5$~kbar, the pressure variation of the superconducting transition temperature is comparable.

Fig.~\ref{Fig_x4_PhaseDiagramsDiscussion}(c) shows the low-temperature phase diagram of  FeSe$_{0.96}$S$_{0.04}$ in an applied magnetic field of 15~T, compared to that of FeSe in 16\,T from~\cite{GuanYu2019}.
Magnetic fields suppress superconductivity and enhance the features associated with
the emergence of magnetism, similar to FeSe under pressure
\cite{GuanYu2019,Sun2016pressure,Kaluarachchi2016}. 
However, the isoelectronic substitution with sulfur affects magnetic interactions and influences
the location of different  magnetic phases.
Interestingly, in high magnetic fields of 15~T,
the maximum in $T_{\rm c}$ for SC1 phase around 7.5~kbar coincides with the onset of the magnetic phase (see Fig.~\ref{Fig_x4_PhaseDiagramsDiscussion}(a)).
However,  at high magnetic fields and pressures, above $p_2$,
the superconducting temperatures of FeSe 
and FeSe$_{0.96}$S$_{0.04}$
are rather similar, despite the reduction in $T_{\rm m}$ of $\sim $ 10~K.
This suggests  that the high-pressure superconductivity of iron-chalcogenide is robust, whereas magnetic phase is more fragile,
similar to that found in Cu doped FeSe \cite{Zajicek2022Cupressure}.
Both $T_{\rm c}$ and $T_{\rm m}$ increase with pressure, suggesting a potential coexistence between the high-pressure
superconducting, SC2, and magnetic phase, M2.

The observed differences in resistivity upturns in high magnetic fields across the pressure phase diagram imply a clear distinction between a rather robust SDW order coupled with orthorhombic distortion inside the nematic phase, called the M1 phase with $C_2$ symmetry ($p_1<p<p_2$), and a more fragile order outside the nematic phase defined as the M2 phase ($p>p_2$) (see Figs~\ref{Fig2_x4_TZ2_rhoxx_vs_T_fixB_fixp}(i)-(l)).
Fig.~\ref{Fig_x4_PhaseDiagramsDiscussion}(d) shows that the pressure dependence of $\Delta_{M}$ (for sample S1) in 15~T closely follows the magnetic temperature, $T_{\rm m}$ of the M1 and M2 magnetic domes but
its energy scale is much larger than $k_{\rm B} T_m$, suggesting that the magnetic field significantly alters the SDW gap and scattering.
Above $p_2$, the resistivity in zero-magnetic field becomes rather linear in temperature and 
increases under pressure, suggesting the change in electronic behaviour inside the M2 phase (see Fig.~\ref{Fig1_x4_zero_field}(d).
This differentiation in magnetism as a function of applied pressure can be
caused by variations in size and lenghtscale of spin fluctuations,  as observed 
in FeSe$_{0.9}$S$_{0.1}$ using NMR studies \cite{Rana2020}.
The magnitude of ($1/T_1T$) AFM spin fluctuations in FeSe$_{1-x}$S$_{x}$ was found to be larger inside the nematic 
phase ($C_2$ phase) as compared {\color{red} to} the high-pressure phase outside the nematic phase ($C_4$ phase),
despite the fact that AFM spin
fluctuations are more effective in enhancing superconductivity in the absence of nematicity. Additionally, the AFM correlation length $\xi_{AFM}$ was estimated to be longer inside the nematic phase \cite{Rana2023}.

The presence of two separate superconductivity domes suggests that the pairing mechanism changes under pressure in FeSe$_{0.96}$S$_{0.04}$ 
(see 
Fig.~\ref{Fig_x4_PhaseDiagramsDiscussion}(a)-(c)).
 Inside the nematic phase, the superconductivity displays a minimum where magnetic phase M1 has a maximum, suggesting a likely competition between the two of them.
The robustness of the high-pressure superconducting phase, SC2, to different substitutions \cite{Zajicek2024,Zajicek2022Cupressure}
indicates either $s_{++}$ sign-preserving pairing or a complex superconducting phase coexisting with fragile magnetism inside the phase M2
 (see Fig.~\ref{Fig1_x4_zero_field}(e)).
Surprisingly, the superconducting critical temperature at the nematic end point is not a maximum close to $p_2$, suggesting that nematic critical fluctuations may be quenched \cite{Reiss2020},
there is an electronic phase coexistence or potential quantum Griffiths phases \cite{Reiss2021}.
Close to $p_2$, the relative widths of the superconducting transitions, 
$\Delta T_{\rm c}/T_{\rm c}$, 
are enhanced (see Fig.~\ref{Fig1_x4_zero_field}(f)) and 
a weak hysteretic behavior can  be detected in magnetic field (see  Fig.~S6 in the SM \cite{SM}).

The nature of magnetic order of FeSe is rather complex and varies with applied pressure.
One one hand, at ambient pressure, the lack of long-range static order is explained due to the competition between N{\'e}el and stripe spin orders \cite{Wang2016b}.
In the presence of applied strain, FeSe is detwined and promotes N{\'e}el  $C_4$ symmetric low-energy magnetic excitations \cite{Chen2019}.
Applied pressure is predicted to change itinerant magnetism by shifting an additional $d_{xy}$ hole pocket at the Fermi level \cite{Yamakawa2017}.
Such a band shift would occur only if the height of the chalcogen increases above the conducting Fe planes, as in the case of the Te isoelectronic substitution \cite{Morfoot2023}.
On the other hand,  $\mu$SR studies in FeSe under pressure postulate that the static magnetic order corresponds to the collinear (single-stripe) antiferromagnetic or bi-collinear order \cite{Khasanov2017}. 
Only local probes using Mossbauer and $\mu$SR spectroscopy detect a small ordered magnetic moment in FeSe of $0.2~\mu_B$  at $p = 40$~kbar \cite{Kothapalli2016,Bendele2010}).
Moreover, the high-pressure superconductivity of FeSe may coexist with magnetic order  \cite{Bendele2012}.

Drastic changes in electronic behaviour and scattering due to the development of novel electronic orders can be assessed from the resistivity behaviour. 
Often in magnetic systems such as BaFe$_2$As$_2$, the resistivity decreases drastically at the onset of the SDW phase,  
as scattering by low-energy magnetic fluctuations
is suppressed due to the opening of a magnetic energy gap
 \cite{Rullier2009,Nakajima2014}.
  Furthermore, the SDW order leads to the Fermi surface reconstruction  induced by nesting between electron and hole pockets \cite{Rotter2008}.
Indeed, a slow quantum oscillation was associated with a Fermi surface reconstruction in the SDW phase of FeSe under pressure \cite{Terashima2014}. 
 The reduction of carrier density, $n$, due to reconstruction consequently has a much larger effect on resistivity which leads to resistivity upturns at low temperature, as observed in FeSe$_{1-x}$S$_x$ (see Fig.~\ref{Fig1_x4_zero_field} and Fig.~\ref{Fig2_x4_TZ2_rhoxx_vs_T_fixB_fixp}) and for Co, K, and P-substituted BaFe$_2$As$_2$ \cite{Rullier2009,Nakajima2014}.

 Changes in magnetoresistance and resistivity upturns reveal the the nature of field-induced electronic phases  as a function of applied pressure, both inside the M1 and M2 phases of FeSe$_{0.96}$S$_{0.04}$.
At high-pressure, 
the magnetoresistance is slowly reduced with increasing pressure
and the magnetic transition temperature, whereas $T_{\rm m}$, and $\Delta_{\rm m}$ activation gap are 
highly  sensitive to magnetic fields (see Fig.~\ref{Fig_x4_PhaseDiagramsDiscussion}(e)).
 Furthermore,  magnetoresistance can be strongly suppressed in highly inhomogeneous phases, or by strong impurity scattering, as in the case of Cu substitution within the conducting Fe plane
\cite{Zajicek2022,Zajicek2022Cupressure}. Thus, the high-pressure phase,
due to the reduction in magnetoresistance and sensitivity of electronic transitions to high-magnetic field (see Fig.~\ref{Fig_x4_PhaseDiagramsDiscussion}(e) and (f)), likely has magnetic regions with reduced correlation length, consistent with NMR studies \cite{Rana2023}.

\paragraph{\bf Conclusions.}

This study gives insight into the complex evolution of magnetic, nematic and superconducting phases in FeSe$_{0.96}$S$_{0.04}$ tuned by pressure.
External magnetic fields reveal large resistivity upturns as signatures of Fermi surface reconstruction and activated behaviour.
The phase diagrams under pressure reveal two different superconducting and magnetic regimes reflecting the competition or coexistence between these electronic phases.
Firstly,  inside the nematic phase, at intermediate pressure, the resistivity upturns are consistent with the development of the SDW phase, and the superconductivity is weakened  where SDW is enhanced suggesting a competition between them. Secondly, at high pressures, resistivity upturns are induced by magnetic field, due to fragile magnetism, which coexist with superconductivity.
The high-pressure superconductivity
remains robust to different chemical substitutions, supporting a sign-preserving $s_{++}$ pairing symmetry.
Our study emphasizes the important role of magnetic fields in inducing and probing electronic orders under applied pressure in unconventional superconductors.

\paragraph{Acknowledgments}

This work was mainly supported by Engineering and Physical Sciences Research Council (EPSRC)
(EP/I004475/1) and Oxford Centre for Applied Superconductivity.
 We also acknowledge financial support of the John
Fell Fund of the Oxford University. 
We acknowledge financial support of Oxford University John Fell Fund.
Z.Z. acknowledges financial support from the EPSRC studentship (EP/N509711/1 and EP/R513295/1).
R.M.A. acknowledges funding from the Margaret Thatcher Scholarship Trust.
I.P. acknowledges funding for an iCASE Studentship (EP/W524311/1) and additional sponsorship
from Oxford Instruments.
P.R.~and A.A.H. acknowledge financial support of the Oxford Quantum Materials Platform Grant (EP/M020517/1). A.A.H. acknowledges support of the Deutsche Forschungsgemeinschaft (DFG; German Research Foundation) under CRC/TRR 288 (Project No. B03).
A.I.C. acknowledges an EPSRC Career Acceleration Fellowship (EP/I004475/1).

\end{document}